# Monitoring Depression in Bipolar Disorder using Circadian Measures from Smartphone Accelerometers

Oliver Carr, Fernando Andreotti, Kate E. A. Saunders, Niclas Palmius, Guy M. Goodwin, and Maarten De Vos*

*Abstract*— Current management of bipolar disorder relies on self-reported questionnaires and interviews with clinicians. The development of objective measures of deteriorating mood may also allow for early interventions to take place to avoid transitions into depressive states. The objective of this study was to use acceleration data recorded from smartphones to predict levels of depression in a population of participants diagnosed with bipolar disorder. Data were collected from 52 participants, with a mean of 37 weeks of acceleration data with a corresponding depression score recorded per participant. Time varying hidden Markov models were used to extract weekly features of activity, sleep and circadian rhythms. Personalised regression achieved mean absolute errors of 1.00±0.57 from a possible scale of 0 to 27 and was able to classify depression with an accuracy of 0.84±0.16. The results demonstrate features derived from smartphone accelerometers are able to provide objective markers of depression. Low barriers for uptake exist due to the widespread use of smartphones, with personalised models able to account for differences in the behaviour of individuals and provide accurate predictions of depression.

*Index Terms*— Bipolar disorder, depression, sleep, activity, hidden Markov models, LASSO regression, mobile-Health.

## I. Introduction

BIPOLAR disorder is a lifelong condition which is characterised by fluctuations in mood states to periods of depression and mania [1]. The World Mental Health survey by the World Health Organisation deemed bipolar disorder to be the illness having the second highest effect on working days lost to illness [2]. With the majority of diagnoses occurring in early adulthood, there is a high cost to society due to bipolar disorder affecting the economically active population [3].

The diagnosis of bipolar disorder is often performed according to a set of guidelines from the fifth edition of the Diagnostic and Statistical Manual of Mental Disorders (DSM-5) [4]. Even in the periods when patients are not experiencing acute episodes, known as euthymia, individuals experience considerable mood instability, presenting a risk factor for a variety of negative psychiatric outcomes [1]. Long-term management of bipolar disorder combines pharmacological and psychological approaches in order to try and prevent transitions into acute episodes [5].

Current long-term management involves monitoring of symptoms between episodes, often performed through clinical interviews and self-reported questionnaires [6], [7]. These methods are subjective and relatively infrequent, thus early detection of deteriorating mood state is difficult and preventative measures are often missed. The development of continuously recorded objective measures to detect deterioration in mood state may considerably improve long-term outcomes by providing early warning signs and allowing preventative measures to take place.

The DSM-5 criteria for diagnosing major depressive episodes in bipolar disorder contains nine sets of symptoms, of which five must be present over a two week period. Some of these symptoms include: reduced interest or pleasure in most activities, insomnia or hypersomnia, and fatigue and loss of energy. A large number of these symptoms for diagnosis of depression are related to sleep and activity levels, therefore in this work we focus on the quantification of objective measures of sleep and activity from accelerometer data with the aim of determining objective markers to aid with monitoring of symptom severity in bipolar disorder.

Previous studies have investigated objective markers for deteriorating mood state in bipolar disorder through both behavioural and physiological measures [8], [9]. A common physiological marker which is often used is heart rate variability [10], with other studies quantifying circadian rhythms through levels of melatonin or cortisol [11], [12]. Circadian rhythms have also been quantified from heart rate and activity recordings, with increased variability and desynchronisation of these rhythms in bipolar disorder patients compared to controls

This work was supported by EPSRC and Wellcome Trust Centre Grants 098461/Z/12/Z (Sleep, Circadian Rhythms and Neuroscience Institute) and 102616/Z (Collaborative Network for Bipolar Research to Improve Outcomes). Oliver Carr acknowledges the support of the RCUK Digital Economy Programme grant number EP/G036861/1 (Oxford Centre for Doctoral Training in Healthcare Innovation).

O. Carr, N. Palmius, F. Andreotti, and *M. De Vos are with the Institute of Biomedical Engineering, University of Oxford, Oxford, UK. K. E. A. Saunders, and G. M. Goodwin are with Department of Psychiatry, University of Oxford, Oxford, UK. M. De Vos is with the Department of Electrical Engineering (ESAT), STADIUS, and the Department of Development and Regeneration, University Hospitals Leuven, Child Neurology, KU Leuven, Leuven, Belgium (correspondence e-mail: maarten.devos@kuleuven.be).

[13]–[15]. Behavioural markers have been investigated through measures including activity, sleep, geographic location, and smartphone or social activity [16]–[18].

Sleep disturbances have long been reported in psychiatric disorders [19]. The most common methods for determining sleep quality are through questionnaires and self-reported sleep and wake times [20], [21]. More recently, automated detection of sleep has been performed from activity data with reliable detection of sleep onset and offset achieved in real-world settings [22], [23]. Quality of sleep can be seen in the criteria for diagnosing depression, and has widely been identified as a crucial factor in determining the deterioration of patients with unipolar depression and bipolar disorder into depressive episodes [24]. The StudentLife study quantified sleep timings, finding significant negative correlations between the duration of sleep and both levels of depression and levels of stress [25].

Studies have used accelerometers to quantify measures of sleep and circadian rhythms in bipolar disorder. However, many of these studies use technology which records high quality signals, thus are only able to record for relatively short periods and rarely capture transitions in mood state. In addition, these studies require participants to wear or carry an additional device, thus reducing the likelihood of the devices being adopted in clinical practice for long term monitoring.

Studies have monitored activity levels over a few days, including Benedetti et al. [26] who monitored 48 hours of activity in 39 bipolar disorder participants through a device worn on the non-dominant wrist with measures of sleep calculated from the actigraphy data. They found patients treated with lithium to have increased activity levels in the evening and having later wake times. Salvatore et al. [27] monitored activity for 72 hours in 36 manic or mixed state bipolar disorder patients and 32 controls. The recovered bipolar disorder patients had advanced acrophase (or peak in activity levels) compared to the controls, increased nocturnal sleep and lower average daily activity, with all results independent of mania ratings, depression ratings and medication.

Other studies have investigated rhythmicity of activity over weeks of recordings. One week of activity was recorded in 42 bipolar disorder patients by Gonzalez et al. [28] through wearing a wrist worn device. They found significant correlations between circadian rhythmicity and manic symptoms in bipolar disorder, however no relationship was found between rhythmicity and depressive symptoms. Krane-Gartiser et al. [29] monitored 24 hour activity patterns of 18 hospitalised bipolar disorder patients with mania, 12 hospitalised bipolar disorder patients with depression and 28 controls. Wrist worn accelerometers were worn for 24 hours. Patients with depression were found to have lower activity levels than the manic patients and the controls, they were also found to have increased variability in activity. Activity levels in mania and depression were lower in the morning compared to the controls, but no differences were found in the evenings.

These studies, in addition to our previous work which found strong correlations between circadian variability of activity and sleep measures with mood instability over a week long period [30], suggest being able to monitor activity over longitudinal periods may provide objective markers for levels of depression in bipolar disorder. A small number of studies have attempted longitudinal monitoring of levels of depression through passive methods such as smartphone accelerometers, however they are often limited by small sample sizes.

A study of 12 bipolar disorder participants in which acceleration was passively recorded through smartphones was able to achieve an average precision/recall value of 66% [31]. In 10 bipolar disorder participants, an accuracy of 70% for state prediction using movement features measured from accelerometers was reported [32]. A further study of smartphone based acceleration in nine bipolar disorder participants was performed, in which correlations were found between the overall activity levels and the patients' mood states [34]. Additionally, stronger correlations were found when the acceleration data was divided into daily segments, such as morning or evening activity, compared to the total activity.

However, other studies have found no associations between activity levels in bipolar disorder and episodes of depression or mania. A 12 month study of 13 participants with bipolar disorder found no relation between smartphone acceleration and clinical symptoms in an inter-patient study [35].

The majority of previous studies have relied on simple measures of activity to quantify circadian rhythms, such as mean activity during specific times of the day or the most active ten hours and least active five hours. Recently, a more advanced method has been developed which attempts to measure circadian rhythms from activity data. Huang et al. [36] implemented a time varying hidden Markov model in order to obtain circadian parameters based on the probabilities of transitioning between active and inactive periods. They calculate a 24-hour profile of activity based on multiple days of accelerometer recordings from which circadian rhythmicity parameters are calculated. They show circadian regularity reduces in cancer patients after chemotherapy treatment.

In this work we present a novel method of combining more traditional measures of circadian activity with measures based on time varying hidden Markov models developed by Huang et al. in order to predict levels of depression using accelerometer data recorded from smartphones. We present longitudinal personalised regression models for patients with bipolar disorder which accurately predict levels of depression provide objective tools to aid with the monitoring of symptom severity in patients with mental health disorders.

## II. Materials and Methods

### A. Dataset Description

The data was collected as part of the Automated Monitoring of Symptom Severity (AMoSS) study at the University of Oxford [37]. Participants gave written informed consent for a protocol approved by the NRES Committee East of England - Norfolk (13/EE/0288), all methods were carried out in accordance with the Code of Ethics of the World Medical Association (Declarations of Helsinki of 1975) for experiments involving humans. The AMoSS study collected behavioral data from participants [38], in addition to self-reported clinically validated questionnaires to determine psychiatric state [39], [40]. Participants diagnosed with bipolar disorder were recruited for the study through previous studies, local advertising and word-of-mouth. All the participants were screened by an experienced psychiatrist using the Structured Clinical Interview for DSM IV. Participants agreed initially for a three-month period of recording, with the option of continuing participating in the study and providing data beyond the first three months. Data collection started in March 2014 and the data analysed here was collected up until January 2018, with 54 bipolar disorder participants recruited.

Participants were given a smartphone (Samsung Galaxy S3 or S4) as part of the study, which they were requested to carry it with them at all times, either as their primary phone or a secondary phone. The smartphone uses a three-directional accelerometer with a sampling rate of 100Hz, a resolution of 0.1mg and is able to record acceleration of positive or negative 2g. The acceleration data is sent to remote servers to allow continuous recordings.

Participants were prompted to complete the clinical questionnaires on a weekly basis by email. The 16 item quick inventory of depressive symptomatology (QIDS) questionnaire was used to determine the levels of depression in participants, assigning a score ranging from 0 to 27, with a score greater than 10 being defined as depression. Table I shows the number of weeks of data recorded from all participants, the number of weeks labelled with a clinical questionnaire score and some participant demographics.

## B. Quantifying Sleep and Circadian Measures

Acceleration signals from the smartphone were segmented into weeklong periods between midnight on Sundays. Total acceleration is used due to variations in orientations in the smartphones while being carried. An example of a raw acceleration signal and the log transform are shown in Figure 2, the raw signal shows the total acceleration, calculated as the square root of the sum of squares of the tri-axial accelerations. Log transforms are used to emphasise smaller accelerations. Depression labels are assigned to each week of acceleration data from any questionnaires provided during that week. If no scores were provided that week, interpolation between the adjacent scores is used if within one week of the record to be labelled, otherwise the week is excluded.

Traditional circadian measures from accelerometer data are calculated from the transformed data, including: the least active five hour period within a day ($L5$) and it's time, the most active ten hour period ($M10$) and it's time, and the relative amplitude $(M10 - L5)/(M10 + L5)$.

Hidden Markov models (HMMs) are used to determine a sequence of unobserved hidden states for an observed sequence, for example periods of activity and rest from an observed accelerometer signal. The models are assumed to be Markov processes in which the current hidden state only depends on the previous state, with probabilities of transitioning between states determined by the model.

A two state HMM is used here to determine periods of activity and inactivity, with the transformed acceleration as the observation sequence. The transition probabilities are allowed to vary with time in order to determine a weekly activity profile [36]. This method, developed by Huang et al., allows the natural circadian cycle of activity in individuals to be captured, in addition to quantification of 24-hour profiles from the varying transition probabilities, thus providing interpretable circadian features for prediction of depression.

There are three main steps to solve for HMMs. Firstly, given the observation sequence $O$, and model parameters $\theta$, calculate the probability of the observation sequence given the model, $P(O|\theta)$. Secondly, optimise the model parameters $\theta$ in order to maximise $P(O|\theta)$. Finally, select the sequence of hidden states $S$ given the observations, $O$, and model parameters, $\theta$. Traditionally the model parameters can be optimised through expectation maximisation in order to find the initial state probabilities, $\pi$, the transition probabilities, $A$, and the observation probabilities, $B$.

Two covariates are introduced to the HMM, a sine wave and a cosine wave with 24 hour periods, making up $X$. This ensures no repeated occurrences of $X$ over 24 hours and introduces a circadian component to the HMM. A multinomial logistic link function is used to calculate the transition probabilities, $a_{i,j}(t)$ from the covariates, given by

$$a_{i,j}(t) = P(S_t = j | S_{t-1} = i, X_t) = \frac{\exp(c^0_{i,j} + c^1_{i,j} X'_t)}{\sum_{j=1}^{m} \exp(c^0_{i,j} + c^1_{i,j} X'_t)}, \quad (1)$$

where $S_t$ is the hidden state at time $t$, $X_t$ is the vector of covariates at time $t$, and $m$ is the number of hidden states. $c^0$ and $c^1$ are vectors of coefficients to be optimised, which represent the relationship between the 24 hour covariates and the probability of transitioning between hidden states.

The joint likelihood of the time varying HMM with covariates $X$ is defined as

$$P(O, Q | \theta, X) = P(S_1 = i | X_1, \pi) \prod_{t=1}^{T-1} P(q_{t+1} = j | q_t = i, X_t, A_{i,j}(t)) \prod_{t=1}^{T} P(O_t | q_t = i, X_t, B), \quad (2)$$

where $T$ is the total length of the sequence and $Q$ is the sequence of hidden states.

An initial selection of model parameters is made, with expectation maximisation iteratively updating the current model parameters, $\theta^*$, to a new set of parameters, $\theta$. With the introduction of covariates, the HMM parameters cannot be optimised with the traditional Baum-Welch algorithm and numerical optimisation is required to estimate the parameters. The expectation of the log-likelihood function can be split into three components, with each component being associated with one of the model parameters: $\pi$, $A$, and $B$.

TABLE I
DEMOGRAPHICS AND ACCELERATION DATA FROM THE SMARTPHONES WITH CORRESPONDING CLINICAL LABELS

| Description | |
|---|---|
| Participants with any labelled weeks | 52 |
| Total weeks labelled | 2105 |
| Weeks per participant (mean ± std) | 37 ± 33 |
| Participants with at least five labelled weeks | 43 |
| Weeks labelled as depressed (<20% missing) | 434 |
| Depressed weeks per participant (mean ± std) | 10 ± 16 |
| Weeks labelled as euthymic (<20% missing) | 1095 |
| Euthymic weeks per participant (mean ± std) | 25 ± 21 |
| Gender | 15 m, 28 f |
| Age (mean ± std) | 38 ± 19 |
| BMI (mean ± std) | 26.8 ± 6.8 |

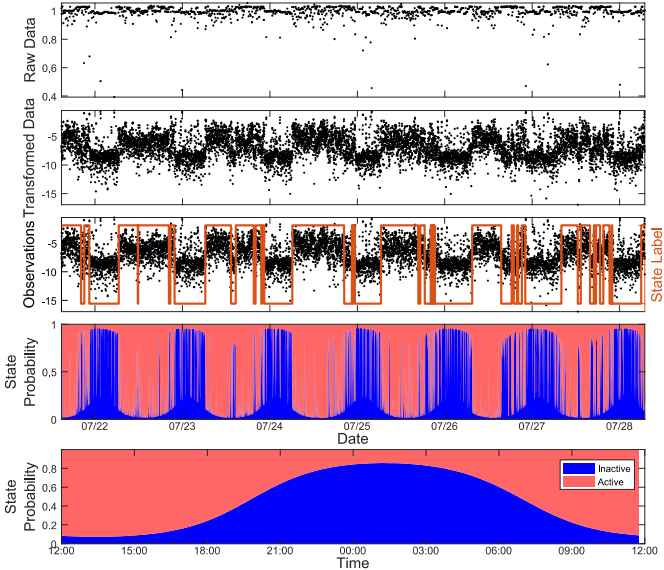

Fig. 2. A week of raw acceleration data recorded from the smartphone for one participant, with log-transformed used as the observation sequence for the HMM. The Viterbi sequence of most likely hidden states, calculated from the HMM, is shown overlaid on the observation sequence. State probabilities are shown

$$Q(\theta,\theta^*) = \sum_{i=1}^{N}\gamma_1(i)\log P(q_1 = j|\pi, X_1)$$
$$+ \sum_{t=1}^{T}\sum_{i=1}^{N}\sum_{j=1}^{N}\xi_t(i,j)\log P(q_{t+1}=j|q_t=i,X_t,A_{i,j}(t))$$
$$+ \sum_{t=1}^{T-1}\sum_{i=1}^{N}\gamma_t(i)\log P(O_t|q_t=i,X_t,B) \quad (3)$$

The parameters $\gamma_t(i)$ and $\xi_t(i,j)$ in (3) describe the probability of being in state $i$ at time $t$, and the probability of being in state $i$ at time $t$ and state $j$ at time $t+1$ respectively, as defined in [41]. As this work only allows the transition probabilities, $A$, to vary with time, the initial states, $\pi$, and the observation probabilities, $B$, can be estimated using the Baum-Welch algorithm.

Combining (1) with the second term in (3) gives
$$f_{max} = \sum_{t=1}^{T}\sum_{i=1}^{N}\sum_{j=1}^{N}\xi_t(i,j)\log\frac{\exp(c_{i,j}^0+c_{i,j}^1 X_t)}{\sum_{k=1}^{m}\exp(c_{i,j}^0+c_{i,j}^1 X_t)}, \quad (4)$$

which is the function to maximise through selection of the coefficients $c^0$ and $c^1$. These coefficients are solved for numerically within each iteration of the Baum-Welch algorithm using an interior-point algorithm. Once the HMM parameters have been optimised, the Viterbi algorithm is used to find the most likely sequence of hidden states.

Figure 2 shows an example of a week of acceleration recordings from the smartphone and the log-transformed data. The sequence of hidden states is also shown in Figure 2 after optimisation of the initial state probabilities, the observation probabilities, and the time varying transition probabilities.

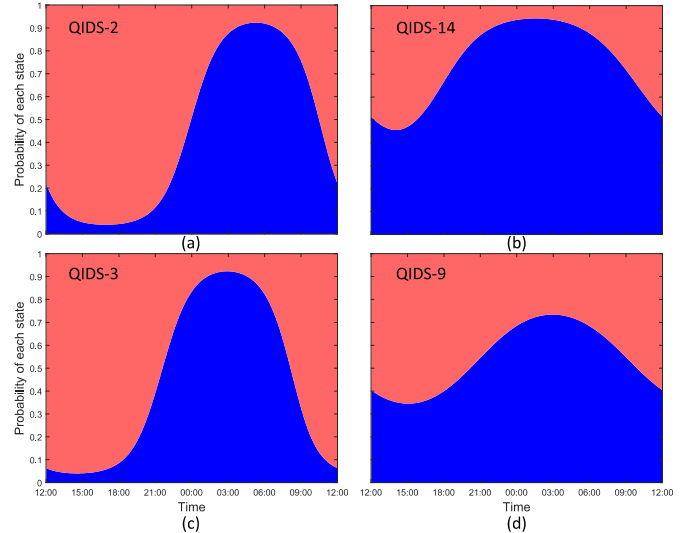

Fig. 3. Four examples of 24-hour profiles derived from the time varying transition probabilities of HMMs fit to a week of acceleration data. The four examples come from different participants show the corresponding QIDS scores of 2, 3, 9 and 14 for depression for that week.

A 24-hour profile is calculated from the time varying transition probabilities of the HMM by
$$P(S_t = j) = \sum_{i=1}^{N} A_{i,j}(t) P(S_{t-1}=i), \quad (5)$$
and is shown in the final plot of Figure 2. Matlab implementation of the time varying HMM can be found at *https://github.com/oliver-carr/Time-Varying-Hidden-Markov-Model*.

*C. Feature extraction*

Features are derived from the Viterbi sequence, including the mean and standard deviation of: a) the estimated sleep time or longest duration in the inactive state, b) the estimated sleep onset and offset or the onset and offset time of the longest period in the inactive state, and c) the total number of times the inactive state is entered. Features are also derived from the 24-hour profile. These include: d) the average inactive period or the time the probability of being inactive is greater than 0.5, e) inactive period onset and offset or the times the probability of being inactive becomes greater or less than 0.5, f) inactive area or the area of the inactive state with probability of being inactive greater than 0.5, g) the maximum probability of being in the inactive state and this time, h) the maximum probability of being in the active state and this time, and i) the rhythm index which is a value between zero and one used to describe circadian regularity as described in [36].

Figure 3 shows four examples of the 24-hour profiles derived from the time varying transition probabilities of an HMM fit to one week of acceleration data from four participants. Figure 3 (a) indicates active behavior during the day and not at night, the steep gradients transitioning between states suggest regular timings of sleep (around midnight) and wake (around 10am), with the week labelled with a low QIDS score of 2. Figure 3 (b) indicates the individual is much less active during the day, the gradients of the profile are not as steep suggesting more variable sleep and wake times, with the week labelled with a high QIDS score of 14. Figure 3 (c) shows a similar profile as in (a) however shifted, with earlier sleep and wake times (approximately 10pm to 9am). This week also is associated with a low depression score of 3. Figure 3 (d) shows a less prominent profile, again associated with variable sleep and wake times, the lower peak also suggests more active behavior at night. This week is associated with a relatively high depression score.

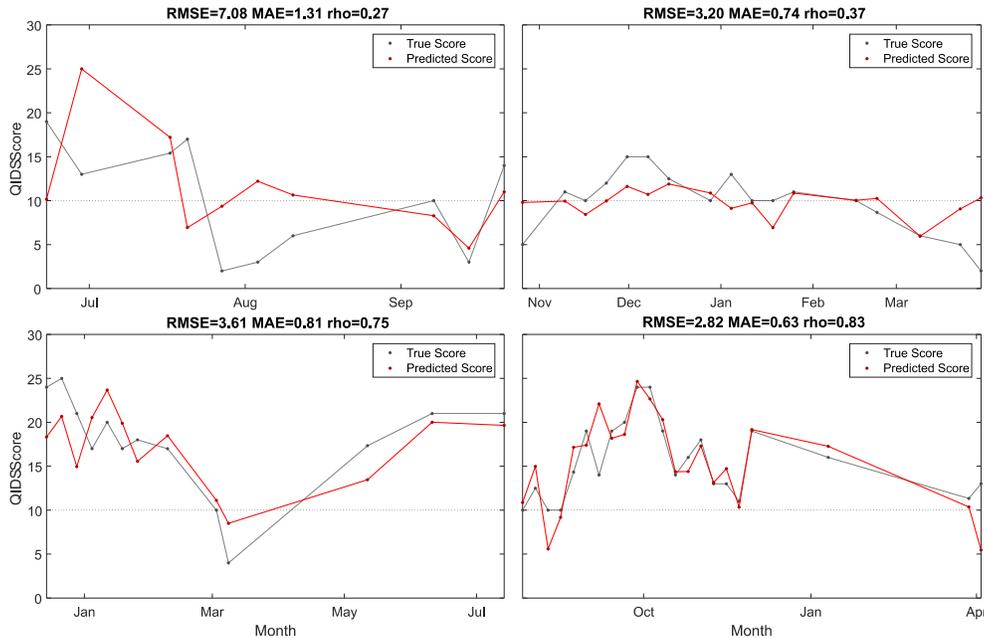

Fig. 4. Four examples of the results of the lasso regression for depression scores from activity derived features from the smartphone. True QIDS scores from the self-reported questionnaires are marked by gray circles, with the predicted scores marked in red. Root mean square error (RMSE), mean absolute error (MAE), and Pearson correlation coefficients (rho) are shown for each of the four participants.

### D. Prediction of Depression Scores

Due to low participant numbers, activity and sleep being so variable between individuals and the depression scores being subjective, it was decided to train prediction models on each individual participant rather than combining their data.

Lasso regularised logistic regression is applied to each individual using two training methods: leaving one week out for testing and training on all remaining weeks and secondly prospective monitoring, leaving the final third of each participant's data for testing and training on the remaining data. A single lasso regression coefficient is chosen for each participant. This is the single value which minimises the mean square error of the prediction across all weeks. Performance of the regressions are quantified through the root mean square error (RMSE), mean absolute error (MAE), and Pearson correlation ($\rho$). The clinical threshold of the QIDS depression score being greater than 10 is used to give each week a binary label of depressed or euthymic, with accuracy, sensitivity and specificity used to measure the performance of the prediction of depression.

## III. RESULTS

Table II shows the mean and standard deviation of the individual lasso regression performance measures for leave one week out regression and prospective monitoring for each participant. The results are also separated into regressions using only the traditional features derived directly from the acceleration data, in addition to regression including features derived from time varying HMMs.

TABLE II
RESULTS OF THE REGRESSION OF QIDS DEPRESSION SCORES FROM SMARTPHONE ACTIVITY MEASURES ON 43 BIPOLAR DISORDER PARTICIPANTS

| Measure | L.O.O. Accel. | L.O.O. All | Last third Accel. | Last third All |
|---|---|---|---|---|
| RMSE | 3.36± 1.61 | 3.28 ± 1.55 | 4.80 ± 3.27 | 4.85 ± 3.19 |
| MAE | 0.91± 0.58 | 1.00 ± 0.57 | 0.48 ± 0.31 | 0.61 ± 0.34 |
| $\rho$ | 0.09± 0.30 | 0.20 ± 0.32 | 0.07 ± 0.40 | 0.14 ± 0.39 |
| N | 4.07± 2.16 | 9.45 ± 6.66 | 5.42 ± 3.69 | 9.28 ± 8.60 |
| Acc | 0.82± 0.19 | 0.84 ± 0.16 | 0.72 ± 0.27 | 0.72 ± 0.27 |
| Sen | 0.41± 0.43 | 0.48 ± 0.43 | 0.13 ± 0.17 | 0.14 ± 0.16 |
| Spec | 0.73± 0.36 | 0.76 ± 0.32 | 0.85 ± 0.26 | 0.86 ± 0.23 |

Results shown as mean ± standard deviation. L.O.O. – leave one out regression, Accel – traditional acceleration features, RMSE – root mean square error, MAE – mean absolute error, $\rho$ – Pearson correlation coefficient, N – number of features used, Acc – accuracy, Sen – sensitivity, Spec – specificity.

Leave one week out regression of QIDS score with only traditional acceleration features achieved a RMSE of 3.36 on a scale from 0 to 27 and a MAE of 0.91. Even with RMSE penalising larger errors heavily, both RMSE and MAE are well within the threshold ranges for diagnosis of the level of depression (0-5 no depression, 6-10 mild depression, 11-15 moderate depression, 16-20 severe depression, and 21-27 very severe depression). The mean correlation coefficient was 0.09, with a mean 4.07 features used. The mean correlation is low due to little variability in QIDS scores of many participants. The classification of depression based on scores over 10 being labelled as depressed achieved a mean accuracy of 82%. The mean specificity is high with a value of 73%, whereas the sensitivity of prediction of depression is lower at 41%.

For leave one out regression using all features, there were small changes in error metrics, however there was an improvement in the correlation between predicted and true scores. The average RMSE of QIDS prediction is 3.28, with an average MAE of 1.00. Correlation coefficients between the true and predicted QIDS scores vary across participants with a mean of 0.20 and a standard deviation of 0.32 due to many participants having little variation in their depression scores, resulting in low correlation coefficients. The number of features used in the lasso regression also varies greatly between participants, with some participants only using two features for the optimal prediction, whereas others require all features to be used. The classification achieved a mean accuracy of 84%. The mean specificity is 76%, whereas the sensitivity is 48%.Prospective monitoring using only past data for training and testing on the final third of the data achieved slightly worse results than leave one out regression, however this is expected due to the reduction in training data and examples of depressed or euthymic episodes [16]. The results using just traditional acceleration features and the inclusion of HMM features were very similar, with RMSE of 4.80 and 4.85, and MAE 0.48 and 0.61 respectively. The MAEs are reduced compared to leave one out regression, this is likely due to lack of variation in the predictions of prospective monitoring due to reduced training data thus often less variability in training QIDS scores. This also reduces the sensitivity in the classification of depression.

Figure 4 shows four examples of the QIDS prediction from smartphone activity features using the leave one week out method. All examples show participants who have true QIDS scores both above and below the threshold to be labelled as depressed, with the predicted results tracking these changes with small errors. The examples show a range of performances of regression results, with an example of high RMSE (7.08) and a correlation coefficient of 0.27, and an example of a relatively low RMSE (2.82), despite the high variation in QIDS scores, and a high correlation coefficient of 0.83.

## IV. Discussion and Conclusion

Monitoring of symptom severity in bipolar disorder currently relies on self-reported questionnaires and interviews with clinicians. This work aims to find objective measures which can be passively recorded to aid with monitoring of depression. We use accelerometers in smartphones to estimate features related to sleep and circadian rhythms in order to predict depression scores from personalised models.

The widespread use of smartphones in modern society allows of continuous and unobtrusive recordings of activity signals through accelerometers, from which measures of activity and sleep may be derived. Only a small number of existing features to quantify circadian rhythms from accelerometers exist in the literature, which are simple measures based on the most active and least active periods in a day. New features have been derived from time varying HMMs which predict periods of rest and activity from the acceleration signals as observations. These features, calculated on a weekly basis, were able to predict depression scores from self-reported QIDS questionnaires with high accuracy (84%) and small errors (MAE of 1.00 from a range of 0 to 27).

Although personalised regression performs well, more data from each individual would likely improve performance. Instances of depression are required in the training data in order for the models to learn from and detect future episodes. For some individuals depression can be relatively rare, thus increasing the time required for data to be collected. In order to reduce this data requirement, yet keep some personalisation in the prediction, clustering techniques could be applied to group participants who behave in similar ways [16].

Although recording activity levels from smartphones provides few barriers for uptake, they still have a number of limitations. Smartphones are limited by their power and need charging, this means they cannot be carried by the participant at all times resulting in missing data. HMMs are able to deal with missing data well, however as smartphones will not be carried during sleep, there will be a period before going to sleep and after waking up where smartphones will not be used. These periods make it difficult to accurately estimate sleep and sleep timings from smartphone data.

The development of smart watches has the potential to improve the quality of continuous accelerometer data from participants for healthcare applications [42]. The standardised location of devices and reduced likelihood of participants not carrying or wearing the device for long periods is likely to improve the detection of active and inactive periods as well as more accurate sleep and wake timings. However, current work comparing activity monitoring from smartphone and smart watch data yields similar results [43].

Further work is required to validate the features used in this work against true values, such as sleep onset and sleep offset times. This lack of validation makes it difficult to interpret the true meaning of the features, however as the regressions are personalised the features are only required to be consistent within individuals. This work uses the activity data from the concurrent week as the depression score, thus providing no warning for deterioration of mental health. Future work should focus on the detection of deteriorating circadian rhythms and sleep measures, which can act as early warning signs of deteriorating mental health. Early signs of depression can prompt early intervention from clinicians to try and prevent transitions into episodes of depression which, once entered, are difficult to revert back to healthy states.

This work has developed novel features to measure circadian variability which can be easily interpreted. This work also demonstrates a strong relationship between circadian rhythms of activity, sleep measures and levels of depression, indicating smartphone activity data may be a potential tool to complement the current management methods in bipolar disorder. The widespread use of smartphones provides an unobtrusive tool for monitoring symptom severity, having the potential to provide early warnings of transitions into depression and allowing for preventative measures to take place.